\documentclass[preprint]{aastex}

\input{psfig}

\shorttitle{Supernova Reverse shocks and SiC Growth}
\shortauthors{Deneault, Heger, Clayton}

\begin{document}


\title{Supernova Reverse Shocks and SiC Growth}


\author{Ethan A.-N. Deneault\altaffilmark{1}, Donald D. Clayton\altaffilmark{2}}
\affil{Department of Physics and Astronomy, Clemson University, 
	Clemson, SC 29634}
\author{Alexander Heger\altaffilmark{3}}
\affil{Department of Astronomy and Astrophysics, University of Chicago,
	Chicago, IL 60637}

\altaffiltext{1}{edeneau@antares.phys.clemson.edu}
\altaffiltext{2}{cdonald@clemson.edu}
\altaffiltext{3}{alex@oddjob.uchicago.edu}


\begin{abstract}
We present new mechanisms by which the isotopic compositions of X-type 
grains of presolar SiC are altered by reverse shocks in Type II supernovae. 
We address three epochs of reverse shocks: pressure wave from the H 
envelope near t = 10$^6$s; reverse shock from the presupernova wind near 
10$^8$-10$^9$s; reverse shock from the ISM near 10$^{10}$s. Using 1-D hydrodynamics 
we show that the first creates a dense shell of Si and C atoms near 10$^6$s 
in which the SiC surely condenses. The second reverse shock causes 
precondensed grains to move rapidly forward through decelerated gas of 
different isotopic composition, during which implantation, sputtering and 
further condensation occur simultaneously. The third reverse shock causes 
only further ion implantation and sputtering, which may affect trace 
element isotopic compositions. Using a 25M$_{\odot}$ supernova model we propose solutions to the following 
unsolved questions: where does SiC condense?; why does SiC condense in 
preference to graphite?; why is condensed SiC $^{28}$Si-rich?; why is O richness 
no obstacle to SiC condensation?; how many atoms of each isotope are 
impacted by a grain that condenses at time t$_0$ at radial coordinate r$_0$? 
These many considerations are put forward as a road map for interpreting 
SiC X grains found in meteorites and their meaning for supernova physics.
\end{abstract}




\section{Introduction}

\begin{figure}[ht]
\centerline{\psfig{figure=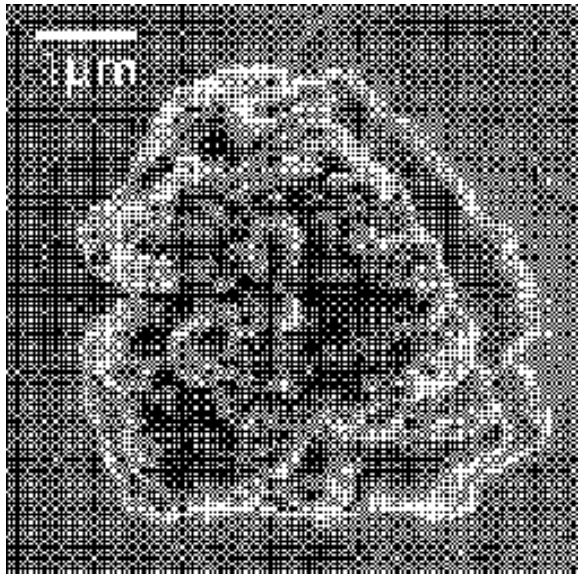,width=3.0in,angle=0}}
\caption{\it SEM photograph of SiC grain KJGM4-244-2 from the Murchison 
meteorite (courtesy of Larry Nittler). $^{12}C/^{13}C = 364 \pm 18$, 
$^{14}N/^{15}N = 82 \pm 3$ in bulk. The highly inhomogeneous appearance of this grain has never been 
explained, although it looks like a collection of smaller grains fused 
together. Such a possibility is predicted in section 6.4 whenever the large 
grains overtake smaller ones following a reverse shock.}
\label{SiCX} 
\end{figure}

\indent Presolar grains which were trapped in the assembly of the
parent bodies of meteorites are extracted and
studied in terrestrial laboratories \citep*{1997ails.conf.....B}. Dramatic isotopic
anomalies (in comparison with solar isotopic abundances) identify these
grains as being presolar and even identify, in many cases, the type of
stellar mass loss within which they condensed. The most thoroughly studied
types of grains are silicon-carbide crystals \citep*[see][for a recent 
discussion of the families of SiC particles]{2001ApJ...559..463A}. In particular, SiC type-X grains
condensed within the interiors of supernovae during their expansion and
cooling \citep*{1992ApJ...394L..43A, 1997Ap&SS.251..355C}. This paper
will concern itself mainly with this SiC supernova condensate (SuNoCon).
We first present arguments for which volume element of the supernova the SiC grains
condense within, and later discuss why SiC can condense at all. We also will
present arguments about why silicon found in SiC is isotopically light in
comparison with solar isotopic composition; and in the same spirit why
isotopically heavy Si in supernova SiC grains is rare.
An SEM photograph of one such X grain (L. Nittler, private communication) 
can be seen in Figure \ref{SiCX}, whose caption illustrates the dramatic 
isotopic anomalies in these grains.\\

\indent A major problem in SuNoCon interpretation has been that
although the grains clearly condensed within supernova interiors, it is not
clear from which parcels of gas the grains condensed. Some authors
\citep*[e.g.,][]{1999ApJ...510..325T} postulate mixing matter from different, 
selected portions of the supernova interior at the molecular level prior
to thermal condensation; but the requisite timescale for this mixing appears
to us to be too long \citep*[e.g.,][]{1999asra.conf..175C,2000LPI....31.1032C}
In addition, such an ad hoc theory does not explain why many other
isotopic compositions that can be obtained from arbitrary mixes are not found in
the laboratory inventory of SiC SuNoCons. We suggest the possibility of
mixing of a different type; namely, grain motion forward at high speed
following the slowing of the gas by reverse shocks. This new proposal was
recently advanced by \citet*{2002ApJ...578L..83C}. We point to importance
of three distinct reverse shocks in Type II supernovae and suggest
their possible cosmochemical importance for the origin of the observed
SiC grains. We study this question anew within the framework of the new
1-D supernova calculations by \citet*{2002ApJ...576..323R}, focusing on their 
25M$_{\odot}$ model s25, which was exploded with a final kinetic energy of
1.2$\times$10$^{51}$ergs. These models are further described by 
\citet*{2002RMP..Inpress}, hereafter WHW.

\section{Reverse Shocks}

At least three reverse shocks are significant for the
condensation of SiC. Our goal is to discuss the implications of these
shocks for the observed SiC X grains from supernovae. Our scientific
results are new ideas and their possible implications rather than a study
of the shocks themselves, for which many uncertainties and
complications exist. We will idealize the shocks as seems befitting for an
initial discussion of their chemical consequences.

\subsection{Reverse shock rebounded from the core/envelope interface}

\begin{figure}[ht]
\centerline{\psfig{figure=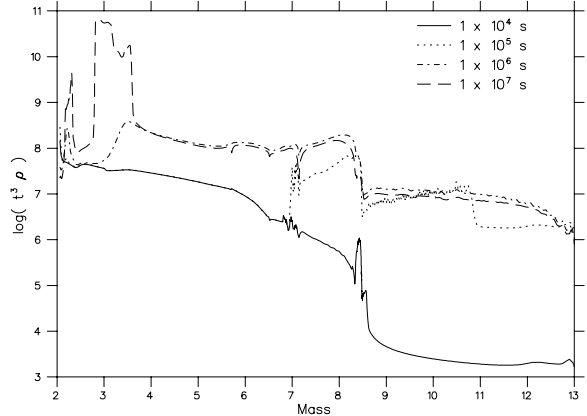,width=3.0in,angle=90}}
\caption{\it The density structure $t^3\,\rho$ versus m at several values of
post-explosion time. Note the large density enhancement in the region
2.7 $<$ m $<$ 3.6. The huge increase at m $>$ 8 for t $<$ 10$^5$s represents the increase of the 
factor t$^3$ prior to the arrival of the shock.} \label{fig1}
\end{figure}

\begin{figure}[ht]
\centerline{\psfig{figure=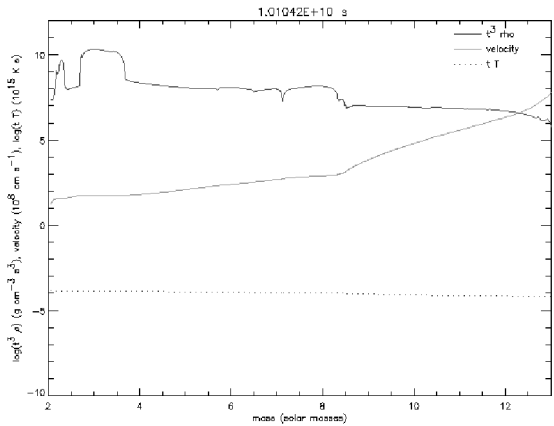,width=3.0in,angle=0}}
\caption{\it An animation which plots density, temperature, and velocity as a function
of mass coordinate. This animation more explicitly reveals the growth of the large density enhancement
found in Figure \ref{fig1}.} \label{movie}
\end{figure}

\begin{figure}[ht]
\centerline{\psfig{figure=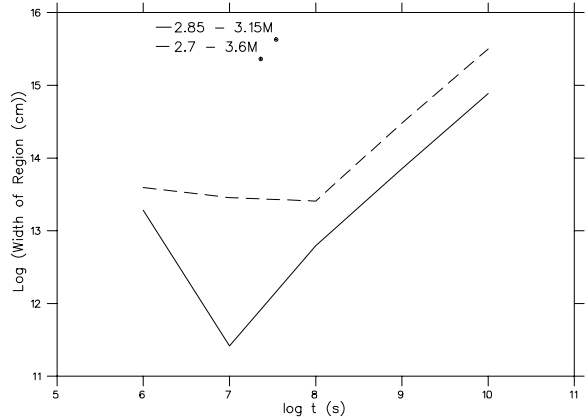,width=3.0in,angle=90}}
\caption{\it The radial width of two mass intervals is shown for selected
values of time. The width of the more dense subshell 2.85 $<$ m $<$ 3.15 declines
by a factor 10$^2$ before 10$^7$ s because its density is held almost
constant over that interval by momentum flux. The larger shell 2.7 
$<$ m $<$ 3.6 reveals less extreme compression by the momentum flux that has
singled out this region in our 1-D calculation. }\label{fig2} 
\end{figure}

\indent This reverse shock is reflected in 1-D hydro from the large
density decrease in the H envelope above the He core. In
10$^6$s it has propagated back inward (in mass coordinate) to $m = 3$ where it establishes a
large peak in the density between $m = 2.7$ and $m = 3.6$. Figure \ref{fig1} plots the product $t^3\rho(t)$
versus radial mass coordinate because that structural product is unchanging 
during epochs of homologous expansion; therefore, it reveals nonhomologous 
expansion as time-dependant structural changes. Figure \ref{fig1} compares 
the density structure at several values of the time to show what can be seen more 
dynamically in an animation. This animation, 
showing density, velocity and temperature versus mass coordinate is Figure \ref{movie} 
In Figure \ref{movie}, the viewer will note that this density peak has not yet been
established at $t = 10^5$s. It is a nonhomologous structure set up by
the interaction of pressure waves, even if the supernova occurs within
a vacuum. There it can be seen that momentum flux toward that 
shell compresses it to higher pressure and density than the surrounding material. The cause of
the inward (in mass) pressure wave that establishes this density peak
is the increase radially of the product $\rho r^3$ in the presupernova structure 
in the hydrogen envelope (see Figure 9 of \citet*{1994ApJ...425..814H} or WHW, Figure 25).
Increasing $\rho r^3$ causes deceleration of the outward shock,
causing the pileup of hot shocked matter that sends the pressure wave
inward in mass coordinate. In space, the trailing material runs into the higher pressure
region caused by the deceleration of the forward shock.\\
\indent The mass interval $2.7 < m < 3.6$ moves outward as a shell of fixed
thickness (approximately) from 10$^6$ to 10$^8$ seconds, confined by the
momentum flux into it, after which it expands homologously from 10$^8$ to
10$^{10}$ seconds. The more dense subshell $2.85 < m < 3.15$ has more extreme
hydrodynamic behavior. Its thickness actually decreases dramatically
between 10$^6$ to 10$^8$ seconds, its density squeezed upward by the momentum
flux into it, after which it expands homologously from 10$^8$ to 10$^{10}$
seconds.  Figure \ref{fig2} shows the radial thickness of both mass intervals
at selected decades of time. For the mass interval $2.7 < m < 3.6$, the near
constancy until $t = 10^8$s is replaced by a linear expansion for larger t.
For the more dense subshell $2.85 < m < 3.15$ the width first decreases sharply
before resuming its participation in a new homologous expansion. The
factor of 100 decrease of the width before 10$^7$s is consistent with the
constancy of the mass contained in a shell whose density remains constant
while the scale of the supernova remnant increases by the factor 10$^2$. This very
dramatic behavior indicates this mass interval as a probable one for dust
condensation.\\

\indent The large density bump will dominate the SiC condensation
structure functions to be described in section \ref{struct}. Not only is this shell more
dense than other supernova matter but its density declines with time
more slowly than t$^{-3}$ during the crucial period for grain growth prior
to 10$^8$s. It may be questioned whether the 1-D hydro code that
prepared this data is adequate for the SiC condensation. \citet{1994ApJ...425..814H} 
showed with 2-D Smooth Particle Hydrodynamics (SPH), that significant
velocity mixing occurs even without an externally launched reverse
shock \citep*[See also][]{2000ApJ...531L.123K}. 
The thin shell in Figure \ref{fig1} will surely not be stable in the face
of reverse shocks, and the question then arises whether the density
enhancement is an artifact of 1-D simulation. For a convincing answer, we must
await 3-D simulations of the reverse shocks within young supernova
ejecta; however the following can be speculated. Although the thin mass
shell may in fact break up into ``clouds'', those may be dense in
comparison with the surroundings for the same reasons that the 1-D
shell is dense, and if so, the dominance of SiC condensation within this
matter may remain. For the time being we emphasize, with this limitation,
the wholly remarkable fact of a dense shell of SiC being established
exactly at the epoch of grain condensation by purely hydrodynamical
interactions. This appears to be an unforseen event, relevant to
physical models of SiC condensation.\\
\indent Within this context it is important to ask about molecular
mixing within the density peak. Is diffusion sufficiently fast
to mix it to a degree that the natural association of $^{29,30}$Si 
with $^{28}$Si is explained by condensation from a mixed gas. We repeat 
\citet{1999asra.conf..175C}'s argument in this specific context to highlight his negative
conclusion. The width of the high density peak at 10$^7$s is about $\Delta r = 10^{12}$cm, 
so the argument consists of showing that diffusion
lengths during a year are very much less than 10$^{12}$cm. With a number
density $n$ = 10$^{12}$cm$^{-3}$ the mean free path $d = 10^4$cm if the
scattering cross section is taken to be 10$^{-16}$cm$^2$. With a temperature
of 2000K while condensation is beginning, the thermal speed is $v_T = 1.6\times
10^5$cm s$^{-1}$ if the mean weight $A = 20$.  Thus an atom scatters 16 times per
second. If its scattered direction is taken to be isotropic, the mean
distance diffused after N scatters is estimated by the isotropic random
walk:
\begin{equation}
	<x^2>^{1/2} = D = d \left( \frac{N}{3} \right) ^{1/2}
\end{equation}
Within a year the number of scatters N at the rate 16 s$^{-1}$ is $N = 5 \times 
10^8$, giving the mean distance $D = 2.2 \times 10^8$cm. This distance 
is much less than the shell thickness $\Delta r = 10^{12}$cm, showing 
that molecular diffusion mixes the matter a negligible amount. Even if 3-D 
calculations confirm, as they surely will, that this thin shell breaks up, the
turbulent ``clouds'' that can be established during a year will still
have sizes greater than 10$^8$cm, so that diffusion is still ineffective
except over these 10$^8$cm interfaces between much larger turbulent
cells. Based on this argument, we hereafter assume that some other
physical mechanism is responsible for causing $^{28}$Si-rich SiC initial
condensates to obtain about half their solar complement of heavy Si
isotopes. We suggest, following \citet{2002ApJ...578L..83C}, that subsequent
reverse shocks must accomplish this. 

\subsection{Reverse shock from massive stellar winds}\label{rev2}

\indent Because of mass loss in the red-supergiant phase, stars
initially more massive than 20M$_{\odot}$ have a large circumstellar shell of
perhaps 1-20 solar masses, depending on the initial mass and on the
mass-loss rates. The new WHW model for a 25M$_{\odot}$ supernova loses 12M$_{\odot}$ 
of material, leaving 13M$_{\odot}$ remaining at the time of explosion.
The 12M$_{\odot}$ of circumstellar material generates a strong reverse 
shock that propagates back, in mass coordinate,
into the nucleosynthesis ejecta on a timescale of months to years,
depending on the structure of the circumstellar material. Spatially,
the ejecta runs into the high pressure created by the shock from the wind.\\
\indent When this reverse shock arrives in the SiC growth region (at
a time between 6 months, when SiC growth begins, and three years, when
SiC growth is very nearly complete) a new mixing mechanism is created for
SiC growth. The condensing grains move forward through the decelerated
gas while SiC condensation is still occurring, sampling a large range
of nuclear compositions. This, in effect, mixes the composition for the
condensing grain. This consideration moves condensation science into
previously unstudied conditions. Namely, the reverse shock heats the
gas to temperatures that are too high for condensation in any
equilibrium sense; but the precondensed grains from the previously cold
flow remain much cooler than the gas. If a grain moves forward at 500 km s$^{-1}$
through the decelerated gas, as we show below, the power input to a 1$\,\mu$m grain owing to 
collisions with atoms at $N = 10^{12}$cm$^{-3}$ is about 10 erg s$^{-1}$. That power
can be radiated by infrared emission at $T_g = 1100$$\,$K. Gas temperatures
of 10$^6$K also can not heat the grains much above 1000$\,$K. Thus the
condensed grains will not evaporate. Indeed, they may even continue
condensation within the hot shocked gas!\\
\begin{figure}[ht]
\centerline{\psfig{figure=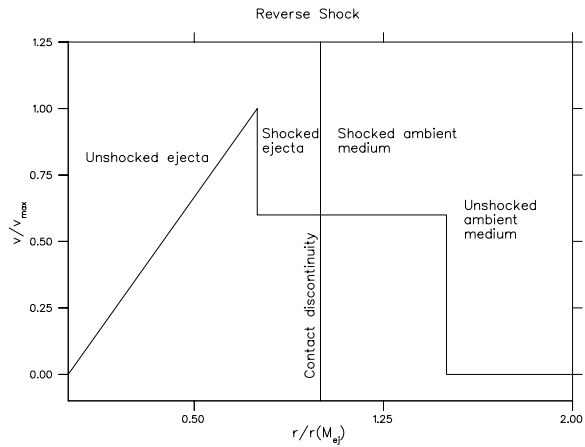,width=3.0in,angle=90}}
\caption{\it Schematic of the velocity profile of the ejecta after a reverse
shock has reentered the ejecta. The velocity increases linearly with
distance (homologous expansion) up to a maximum V$_{max}$ at the radius
reached by the reverse shock. Note that the radial distance is measured
by a mass coordinate labeling that radius, where unity is the contact
surface between ejecta and circumstellar material. }\label{fig3} 
\end{figure}

\begin{figure}[ht]
\centerline{\psfig{figure=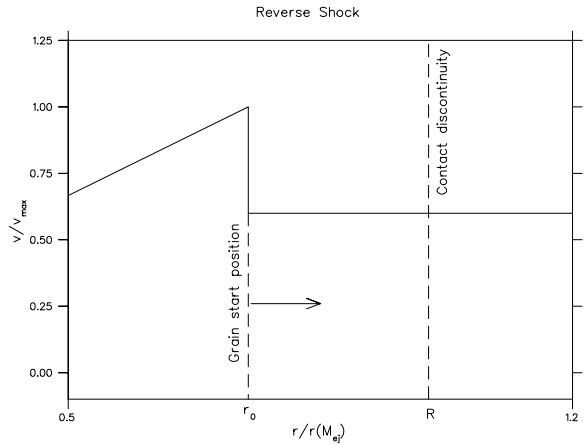,width=3.0in,angle=90}}
\caption{\it Schematic of the relative velocity forward of already condensed
grains after the reverse shock has slowed the flow to about 0.6V$_{max}$
(see Figure \ref{fig3}). The grains move forward with relative velocity 0.4 V$_{max}$
into the already shocked ejecta.}\label{shel} 
\end{figure}

\indent Figure \ref{fig3} is a schematic depiction of the velocity profile
established by reverse shocks \citep*[see also][]{1999ApJS..120..299T}. Velocity
increases linearly with radial distance within the ejecta that has not yet been
shocked by collision with the wind to
a maximimum $V_{max}$ at the location of the reverse shock. External to
that shock, the already shocked ejecta has assumed a nearly constant
velocity within the remainder of the ejecta. That velocity is about 60$\,$\%
of $V_{max}$, and as the shock propogrates inward (in mass coordinate, not
in space), the value of $V_{max}$ declines according to Figure \ref{fig3} and the
external velocity correspondingly declines to maintain a constant
velocity near 60$\,$\% of $V_{max}$. Despite this being an approximation, we
assume it to enable discussion of the cosmochemical consequences.
Notice that the already shocked ejecta is moving outward as a shell
(uniform velocity) whereas the unshocked ejecta expands homologously
until the shock reaches it.\\
\indent Figure \ref{shel} illustrates the situation for a thermally condensed
grain within the ejecta as the shock reaches it. Hydrodynamic forces
slow the gas to about 0.60$\,V_{max}$, but the grain can not be slowed so
rapidly; hence it moves forward with a relative speed $\Delta V = 0.40$
$V_{max}$. If the dense shell $2.7 < m < 3.6$ moves near 1500 km s$^{-1}$, for
example, the grain will move forward through the hot shocked gas at
speed $\Delta V = 600$ km s$^{-1}$. At such speeds atoms striking the grain will
be implanted to depths near 0.05 micron within the grain, and will also
cause sputtering from the grain surface. The grain temperature, however, will
remain near 1000$\,$K, so evaporation will not occur. Only sputtering
can remove atoms at this time. This relative speed is about 50$\,$\% greater
than the familar speed of the solar wind, kinetic energy near 1-2 keV
per nucleon. A large literature studied the sputtering and implantation
of solar wind into lunar fine soil during the Apollo program 
\citep*{1978LPSC....9.1667Z, 1978LPSC....9.1655K}. Because the sputtering yield is near
unity, the grain will only slowly gain or lose mass; but the identity
of the atoms within it may evolve. This conclusion has already been
presented \citet{2002ApJ...578L..83C}. Although the photonic emission from
these shocks has been well studied \citep*{1994ApJ...420..268C,2002ApJ...572..350F}, 
their chemical implications for thermal condensates has not.

\subsection{Reverse shock from interstellar medium}

\indent A huge literature exists \citep{1999ApJS..120..299T} for this
interaction, both for the X ray emission from the shocked ambient
medium and for the emission from the reheated ejecta by the reverse
shock. If the ISM is as dense as compact HII regions, this reverse
shock will reenter the ejecta, and will decelerate the gas with respect 
to the high-inertia SiC grains, which will have already condensed.
This will cause sputtering and ion implantation as the grains
then propagate outward through the slowed gas, as was originally
envisioned by \citet{2002ApJ...578L..83C} within this context of a reverse
shock from the ISM. At these late times, the grains may have actually
emerged from the supernova interior. They will be warm enough for IR
radiation but their interactions will consist entirely of sputtering and
ion implantation, which occurrs throughout their outward journey
through the overlying column of atoms.

\section{The SiC condensation-structure functions}\label{struct}

\indent The condensation of SiC is not as well understood kinematically
as is the condensation of graphite. For graphite, it was possible to construct
\citep*{1999Sci...283.1290C,2001ApJ...562..480C} an
explict kinematic model: stationary molecular abundances of linear
chains C$_n$; isomerization of linear C$_n$ into ringed C$_n$, which is more
resitive to oxidation; rapid attachments of C atoms to rings, which
graphetize during the process. The condensation of SiC will occur in
gas that is not only heavily oxidizing, but in which the relative
abundances of Si and C vary rapidly with location. Using
the mass fractions of Si and of C as function of radial mass in 
25M$_{\odot}$ model by \citet{2002ApJ...576..323R}, the SiC structure
functions to be defined below indicate that Si will be more abundant
than C in the dense shell that is established near m = 3, and that SiC
will condense there.\\
\indent Without a full theory of dynamic SiC condensation, we can
nonetheless motivate where in the supernova that it should occur most
prolifically. We do this with densities of the reacting species Si and C
near 10$^7$s, as in Figure \ref{fig1}.

\subsection{SiC molecule-formation-rate structure function}

\begin{figure}[ht]
\centerline{\psfig{figure=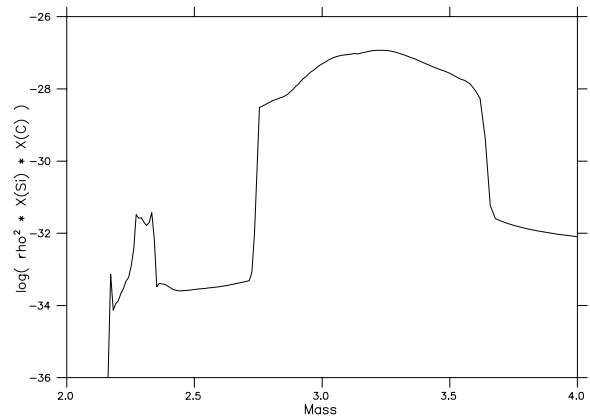,width=3.0in,angle=90}}
\caption{\it First SiC structure function.
The product n(C)n(Si) cm$^{-6}$ which provides the
source term for SiC molecules. One expects SiC solids to grow
in regions where SiC molecules are rapidly made. } \label{fig5} 
\end{figure}

\indent Figure \ref{fig5} shows the product n(C)n(Si) cm$^{-6}$ which provides the
source term for SiC molecules. One expects SiC solids to grow
in regions where SiC molecules are rapidly made. This figure
shows a huge bump between m = 2.7 and m = 3.6. As described below, silicon
is light ($^{28}$Si-rich) in the inner half and heavy ($^{29,30}$Si-rich) 
in the outer half. The number density of Si atoms in this density peak 
at t = 10$^7$s is n(Si)=10$^{11}$cm$^{-3}$, several magnitudes greater 
than in surrounding material. We therefore take it that SiC X grains must condense here.\\

\indent We follow \citet{1999Sci...283.1290C,2001ApJ...562..480C} in assuming
that the buildup of CO and SiO molecules is counteracted by radiative disruption, enabling SiC 
solids to condense even in regions where O $>$ C. The reader must not be disturbed at the thought of
disruption of SiC by the same process, because SiC can react rapidly by
chemical means whereas CO can not. This is a fundamental distinction.
Small SiC crystals are destroyed (in part) by their chemical growth to larger SiC
structures. Thus CO and SiO destruction restores free C and Si to the
gas (See \citet{2001ApJ...562..480C} for a detailed pathway in
the case of graphite condensation). Our supposition that SiC solids
condense quite well in O-rich radioactive gas rests on the idea that
when a condensation nucleus already exits (SiC), it condenses further Si and
C from the free atoms in the gas faster than the free O atoms
can destroy SiC solids. To be sure, O collisions with the grain are
more frequent, but the oxidation of SiC solid is slower because of the
additional need for the O atoms to break the SiC bonds to remove an integrated
atom from a small crystal. The Si and C atoms, on the other hand, can
simply bond and stick. \citet{2001ApJ...562..480C} discussed
this transition to slower oxidation rates for the graphite case, and we
shall assume that the SiC case is similar. Condensation in an oxydizing gas
was not believed to be possible by \citet{1999ApJ...510..325T}, who hypothesized that 
$^{28}$Si-rich inner-core silicon could mix at the molecular level with
enough C-rich gas from the He shell to condense SiC in C-rich mixtures.\\ 
\indent A smaller peak can be found between m = 2.2 and 2.4 M$_{\odot}$, 
in a region of pure $^{28}$Si, but as will be shown in the next section, its
contribution to the formation of SiC grains is negligible. 

\subsection{SiC maximal-growth structure function}

\begin{figure}[ht]
\centerline{\psfig{figure=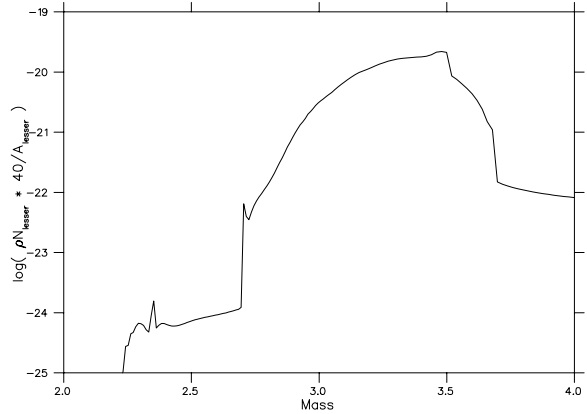,width=3.0in,angle=90}}
\caption{\it The maximum mass density of SiC given by the concentration of
the lesser of the two constituents.}\label{fig6} 
\end{figure}

\indent Figure \ref{fig6} shows a second SiC structure function. It presumes that
condensation of SiC is so efficient that it completely depletes the element of
lesser abundance. In other words, condensation of SiC depletes all Si wherever C is more
abundant and all C wherever Si is more abundant. Figure \ref{fig6}, in that sense is a graph of the maximum
mass density of SiC. It is the product of the supernova density $\rho$ with
the mass fraction of the lesser of the two elemental mass fractions,
augmented by the factor 40/A$_{lesser}$, where A$_{lesser}$ = 12 if C is less
abundant and 28 if Si is less abundant. This product represents the
maximum density of SiC that could possibly condense at each mass
coordinate:
\begin{equation}
\rho(SiC)= \rho\,X_{lesser}(C,O)\,\frac{40}{A_{lesser}}
\end{equation}

\indent This shows that the mass zone of
enhanced density, 2.7 $<$ m $<$ 3.6, also has the greatest capacity to
condense SiC mass when the less abundant species is totally depleted in
SiC formation. The smaller peak, found between 2.2 and 2.4M$_{\odot}$ in Figure \ref{fig5}
has no such capacity for grain formation, even though the rate in that 
region is comparable. The integrated number of collisions is adequate to
deplete condensibles if nucleation sites are able to form early.
\citet{1999Sci...283.1290C} were able to give such a kinetic theory
of nucleation for graphite, but such a compelling picture does not yet
exist for SiC.  Without adequate population control \citep{2001ApJ...562..480C} 
these nucleation sites can be too abundant, in which case
the depletion is total but results in large numbers of tiny particles
rather than the micron-sized examples found abundantly in the
meteorites.\\
\indent Faced with these uncertainties we will in what follows assume
that the depletion of lesser abundance is total but that its numbers
are sufficiently restricted to allow each grain to grow large. That is what
the experimental data for micron-sized SiC demonstrates. 

\section{Why do supernovae condense SiC at all?}\label{condense}

\indent Supernova SiC X grains are rather common, comprising about one percent 
of the number of SiC grains condensed in C-star winds. Several authors
\citep{1996ApJ...472..760B,1995GeCoA...59..1633} have shown that in C
stars graphite will condense prior to SiC unless the C/O ratio is no
more than just slightly greater than unity. (See Figure 10 of \citeauthor{1996ApJ...472..760B}).  
This restriction on the C/O ratio derives, however, from the
assumption that CO formation will lock up available carbon, in which
case, if C/O only slightly exceeds unity, the free C is no more
abundant than is Si, with the result that SiC can condense.\\
\indent But in supernovae, the CO molecule can not permanently bind up
the C atoms. Given solar abundances, graphite will then condense prior to
SiC and deplete the C. If graphite condenses first, the question arises
``Why, then, do supernovae condense SiC?''. This simple question appears not to
have been clearly answered in the literature; therefore we shall address it
here. SiC condenses in preference to graphite when the free Si
abundance exceeds by a factor ten or more the free carbon abundance. In
such circumstances, not only does the equilibrium favor SiC
condensation, but any graphite nuclei that do grow will be rapidly
converted to SiC by the much more numerous Si atoms. It is important in this regard to 
remember that the grains will be cooler than the gas, even during the condensation epoch, 
owing to their infrared cooling.\\
\begin{figure}[ht]
\centerline{\psfig{figure=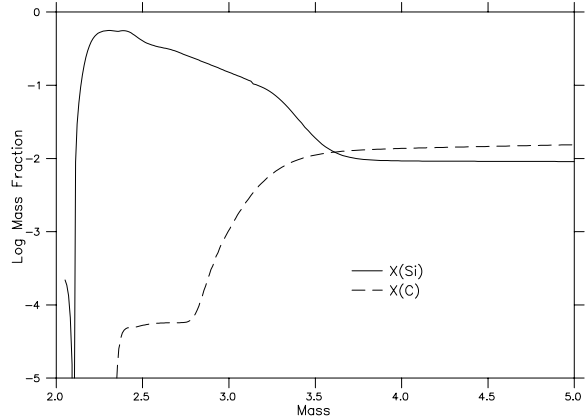,width=3.0in,angle=90}}
\caption{\it Mass fractions of Si and C in the 25M$_{\odot}$ supernova (WHW). Graphite is
expected to condense first, depleting C whenever they are comparably
abundant. To condense SiC requires Si $>$ C. Accordingly we focus on the
region inside m = 3.2, where Si/C $>$ 10, as the supernova portion wherein
SiC is a major condensate.}\label{fig7}
\end{figure}

\indent Figure \ref{fig7} shows the mass fractions of Si and C in the 25M$_{\odot}$
supernova \citep{2002ApJ...576..323R}. Interior to m = 3.4 the abundance of Si exceeds that of
C; and within m = 3.2 by an order of magnitude and more. This region
lies squarely within the density peak (2.7 $<$ m $<$ 3.6) shown in Figure 1\\
\indent It would be misleading to simply apply equilibrium condensation
theory to this situation because the condensation has kinetic controls
that cause the condensates to differ slightly from equilibrium; and
presenting condensation theory \citep*[e.g.,][]{1995GeCoA...59..1633} is not our
purpose. Our present goal is met by the simple assumption that SiC
condenses wherever Si is at least tenfold more abundant that C. Figure \ref{fig7}
then targets the zone 2.7 $<$ m $<$ 3.2. If a Si/C ratio of 100 is instead
required, this same argument targets the zone 2.7 $<$ m $<$ 3.0 instead.
Accordingly, we postulate that SiC SuNoCons form in this matter only.
At larger m, graphite depletes the carbon before SiC condensation can
occur. 

\section{Why are supernova SiC grains $^{28}$Si-rich?}

\begin{figure}[ht]
\centerline{\psfig{figure=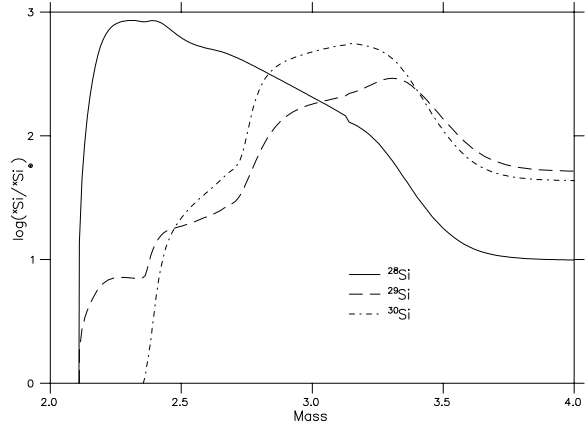,width=3.0in,angle=90}}
\caption{\it Overabundances of the Si isotopes in the 25M$_{\odot}$ supernova with 
solar initial abundances. The
$^{28}$Si-rich portions are inside m = 2.8. If presolar supernovae had Z = 0.5
solar rather than solar, the $^{28}$Si-rich zone would extend to about m =
3.0. Bulk Si/C is much greater than unity within m = 3.0.} \label{fig8}
\end{figure}

\indent The $^{28}$Si excess within SiC X
grains is one of the defining characteristics of supernova SiC. But why?
It is surely not adequate to simply say that ``Supernovae make a lot of $^{28}$Si-rich
Si'', because, as Figure \ref{fig8} shows, they also make a lot of $^{29}$Si and
$^{30}$Si-rich silicon. Where, then, are the $^{29,30}$Si-rich SiC X grains? 
We now believe that an answer to this question also can be given.\\

\indent Based on the results of section \ref{condense}, we anticipate that SiC condenses
(in the 25M$_{\odot}$ supernova) within the zone 2.7 $<$ m $<$ 3.0-3.2. When the
integrals over the mass in that region are performed, the data leading
to Figure \ref{fig8} confirm that the $^{29}$Si/$^{28}$Si ratio is between 1/3 
to 1/2 solar and the $^{30}$Si/$^{28}$Si ratio is slightly in excess 
of solar for stars having initial solar abundances. However, the presolar 
supernovae responsible for the X grains exploded more than 5 Gyr ago. 
Considering also the excess metal richness of the sun for its location 
in the galaxy we can conclude that the average presolar supernova would have 
been perhaps half solar in its initial metallicity. In this case the 
entire Si-rich zone is also $^{28}$Si-rich in its Si. The caption of Figure \ref{SiCX} describes
the isotopic signature of that grain, which is typical for this class of SiC X grains. It is for this reason 
that SiC X grains are $^{28}$Si-rich. The matter outside m = 3.0 - 3.2M$_{\odot}$ condenses 
graphite rather than SiC, as shown in section \ref{condense}, therefore we see 
why the $^{29,30}$Si-rich SiC X grains do not exist.\\
\begin{figure}[ht]
\centerline{\psfig{figure=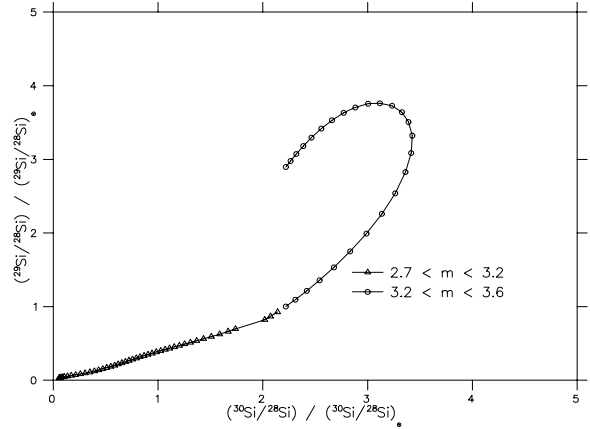,width=3.0in,angle=90}}
\caption{\it Three-isotope plot for Si isotopes in the condensation zone. The
calculated ratios for a supernova of half-solar inital metallicity are normalized 
to the solar value, which is what the laboratory X
grains are compared with. The sequence of triangles traces the Si
isotopic composition from m = 2.7 to m = 3.2; and the circles trace it
through the remainder (3.2 $<$ m$<$ 3.6) of the density peak of Figure \ref{fig1}.} \label{fig9}
\end{figure}

\indent This is demonstrated in Figure \ref{fig9} for a presolar supernova of half
solar metallicity. That is, the supernova $^{29}$Si/$^{28}$Si and 
$^{30}$Si/$^{28}$Si ratios are, in each mass shell m, taken to be half 
of the values calculated in the solar-metallicity 25M$_{\odot}$ model by 
WHW. \citet*{1996ApJ...472..723T} demonstrated that such a procedure obtains reliable results.
The formatting of the plotted points, shell by shell, uses triangles inside m
= 3.2 and circles outside. They make it evident that the normalized
ratios are largely less than unity within m = 3.2 and become greater than unity
outside. If the entire high-density zone 2.7 $<$ m $<$ 3.6 is summed (as if
it were mixed), the half-solar metallicity model gives a bulk Si
characterized by $^{29}$Si/$^{28}$Si = 0.46 solar and $^{30}$Si/$^{28}$Si = 
0.95 solar. With the restriction that only the zone 2.7 $<$ m $<$ 3.2 is able to 
condense SiC, the averages for the half-solar model become  
$^{29}$Si/$^{28}$Si = 0.26 solar and $^{30}$Si/$^{28}$Si = 0.76 solar. Table \ref{massratiotab} tabulates 
these mass ratios for three mass zones: 2.7M$_{\odot}$ - 3.0M$_{\odot}$, 3.2M$_{\odot}$, 
and 3.6M$_{\odot}$, respectively. 
They are even more $^{29}$Si and $^{30}$Si-deficient should SiC be able to condense 
only if the Si/C ratio is greater than 100, as discussed above. \\
\indent The arguments presented here and in section \ref{condense} have shown why SiC condenses in
preference to graphite in certain portions of the supernova and why that
portion is isotopically light. These are the characteristics of the SiC
X grains analyzed isotopically by secondary-ion mass spectrometry. We
submit these ideas as a tentative road map for undestanding these incredible
portions of various presolar supernovae. In the next section, we return to other
consequences of the reverse shocks. 

\section{Column Densities of Overlying Atoms}

\indent Suppose a SiC grain condenses at radial mass coordinate m (which
is comoving with the SN structure). When a reverse shock arrives at m, 
the gas ejecta will be slowed, and the grain moves through the 
overlying ejecta \citep{2002ApJ...578L..83C}. The grains have higher inertia
than the gas ejecta, and so we shall assume that their velocity remains
unaffected, therefore, their relative speed
will equal the amount by which the gas has been decelerated. \citeauthor{2002ApJ...578L..83C}
estimated that drift speed to be near 500km/s for reverse shocks
in the Si-rich zones that do not arrive until 300 yr, as perhaps in Cas
A; but that speed may be comparable if the ejecta encounter a massive
circumstellar wind much earlier, say near 1 yr.\\
\indent \citet{2002ApJ...578L..83C} argued that the grain will encounter the
entire overlying column of atoms within the ejecta and that they will be
implanted within the grains. If so, the number of atoms implanted in the grains is a function 
of the time when the reverse shock reaches the condensation region of the grain. Because the grain
cannot slow until it has encountered a gas mass comparable to its own
mass, it leaves the ejecta at almost that same relative speed if the
reverse shock is at 300 yr, but it will decelerate while still within
the ejecta if the reverse shock arrives within a few years. To see this,
we compute the column densities to which drifting grains will
be exposed. We deemphasize the calculation of shock physics here and of
the slowing of the grains in order to first emphasize the possible
isotopic consequences of ion implantation.

\subsection{Column density upper limit}

\indent All of our calculations were done using model s25 of \citet{2002ApJ...576..323R}. Using this model, 
we can integrate the column densities of all isotopes, starting from the grain's 
condensation radius $r_0$ through the overlying ejecta. SiC grains are observed to have
radii of approximately one micron, so we assume that the cross sectional area of the column 
is $\sigma$ = 1$\,\mu$m$^2$. Then it is possible to compute the number of atoms
that the grain will interact with by constructing a tube of constant 
cross section which passes through all the ejecta overlying the zone 
in which the grain formed. \\
\indent For species $^AZ$ with mass fraction X, the total number of atoms in the column
is given by:
\begin{equation}
N(^AZ) = \sigma \int^R_{r_0} X(r)\,\rho(r)\,\frac{N_A}{\mu_{^AZ}}\,dr\label{naz}
\end{equation}
Where $R$ is any radius larger than $r_0$, but less than or equal to that of the contact 
discontinuity, $N_A$ is Avogadro's number, and $\mu_{^AZ}$ is the atomic weight of the species.\\
\indent 
Another useful form of the integral depends on the Lagrangian mass coordinate, m.
Since $dm$ = $4\pi\,r^2\,\rho(r)\,dr$, and therefore, we can rewrite the integral:
\begin{equation}
N(^AZ) = \frac{N_A\,\sigma}{\mu_Z}\,\int^M_{m_0} \frac{X(m)\,dm}{4\pi r^2(m)}
\end{equation}
Where $m_0$ and $M$ are the grain's starting mass coordinate and the mass 
coordinate of the total ejecta, respectively.\\
\indent It is important to note that both of these integrals are have taken over a static structure. That is, they 
count all of the the atoms in the column at the time that the reverse shock 
arrives at the coordinate at which the grain condenses, and starts 
to move through the ejecta, and does not take into account the time dependent
expansion of the ejecta. The entries in Table \ref{statictab} can therefore be thought of as upper limits to the number
of possible interactions of the grain with selected long lived isotopes of Si, Fe and Mo. 
Table \ref{statictab} is calculated for a shock time of 10$^9$s, and $R$ is the radius corresponding to the 
contact discontinuity. The numbers in the static column density depend on the shock time $t_0$ as (10$^9$s/$t_0$)$^2$. 

\subsection{Column density in a shell}

\begin{figure}[ht]
\centerline{\psfig{figure=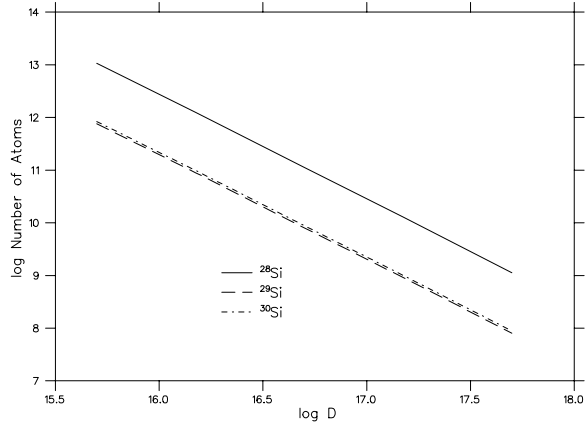,width=3.0in,angle=90}}
\caption{\it Number of Si atoms that a grain can encounter, starting from 
m(r$_0$) = 2.6M$_{\odot}$ as a function of
the distance parameter D = t$_0$$\Delta$V. As $\Delta$V is a constant, 
the number of atoms decreases as t$_0^{-2}$}\label{dparam} 
\end{figure}

\begin{figure}[ht]
\centerline{\psfig{figure=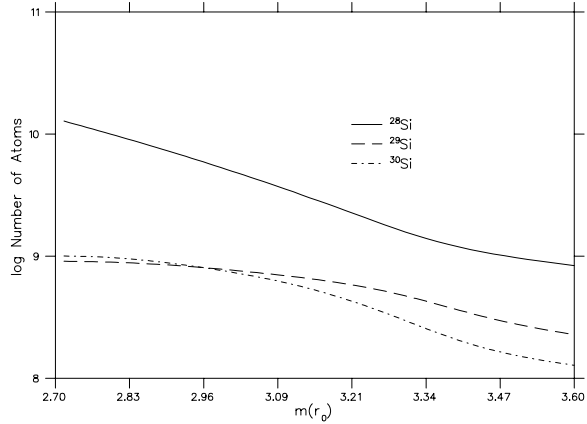,width=3.0in,angle=90}}
\caption{\it Number of Si atoms that a grain can encounter 
as a function of the starting position of the grain in mass coordinate, calculated
for a reverse shock at t$_0$ = 3$\times$10$^9$s.}\label{rparam} 
\end{figure}

\indent As was described earlier in section \ref{rev2}, the circumstellar reverse shock sets up a 
shell structure in the ejecta. We can use this feature of the ejecta to 
track the grain's motion through a constant-width region of decreasing density owing to the
continued expansion of the ejecta. We define r$_0$ to be the radial position where the grain condenses, and R 
to be the outer radius of the shell. For the purposes of our calculation below, 
we assume as before, that R is the radius corresponding to the interface between the He shell and the H envelope,
though R can again be any radius up to that of the contact 
discontinuity. We construct a shell of width $R - r_0$ = constant 
which is moving with the ejecta with velocity $V = 0.6 V_{max}$ (see Figure \ref{shel}). 
We have assumed, as before, the velocity of the grain to be mostly unaffected by 
decrease in velocity of the gas due to the reverse shock, so the grain moves 
through the shell at $\Delta V = 0.4 V_{max}$ = 500$\,$km s$^{-1}$ Since the 
shell's width is constant in $r$, the density in this shell decreases as $t^{-2}$.\\
\indent One may ask whether it is physical to assume that the shell velocity will remain 
uniform over time. Looking at Figure \ref{shel} we see that 
the reverse shock sets up the entire shell to have the same velocity. If we consider 
a small $\Delta t$, $V_{max}$ will decrease, as will the velocity of the shell, however, 
the shell will retain a uniform velocity. Therefore, we can consider $R - r_0$ = constant at all times 
in the shell.\\
\indent At some radius $r^{\prime}$ 
($r_0$ $<$ $r^{\prime}$ $<$ $R$) at some time $t^{\prime}$, the density can be written:
\begin{eqnarray}
\rho(r^{\prime},t^{\prime}) & = & \left( \frac{t_0}{t} \right)^2 \rho(r^{\prime},t_0) \nonumber\\
& = & \left( \frac{t_0}{t_0+\Delta t} \right)^2 \rho(r^{\prime},t_0) \label{rhort}
\end{eqnarray}
Since $\Delta$ t = (r$^{\prime}$ - r$_0$) $\Delta$V$^{-1}$, we can re-write the time dependant term as:
\begin{equation}
\left( \frac{t_0}{t_0+\Delta t} \right)^2 = \left( \frac{t_0 \Delta V}{t_0 \Delta V + (r^{\prime} - r_0)} \right)^2
\end{equation}
We define t$_0$ $\Delta$V = D, where D becomes a distance parameter. Now, we can write an 
integral over dr$^{\prime}$ similar to equation \ref{naz}:
\begin{eqnarray}
N^{\prime}(^AZ) & = & \sigma\int^R_{r_0} X(r)\,\rho(r)\,\frac{N_A}{\mu_Z}\,\left( \frac{D}{D + (r^{\prime} - r_0)} \right)^2\, dr^{\prime} \nonumber \\
& = & \sigma\int^R_{r_0} n(X)\,\left( \frac{D}{D + (r^{\prime} - r_0)} \right)^2 dr^{\prime}
\end{eqnarray}
The distance parameter, $D$, mitigates the number of atoms that the grain 
can interact with. If we assume a constant velocity of the grain through the shell, 
then a one order of magnitude change in D changes the abundance of each atomic 
species by a factor 100, as shown in Figure \ref{dparam}. This was not unexpected,
as the density of the shell decreases as $t^{-2}$. Figure \ref{rparam} shows the 
dependence of the results on the initial position of the grain, $r_0$. This figure 
shows a very important effect of the variable density of the shell. Since, as the 
grain moves forward into its overlying column, the density of the column is 
decreasing, the region that is closest to where the grain condenses has a greater 
effect on the grain's composition than regions farther away. As can been seen in 
Figure \ref{rparam}, grains that form between 3.2 - 3.6 M$_{\odot}$ interact with more 
$^{29}$Si than $^{30}$Si. If grains are aggregations of smaller grains (section \ref{agg}), 
the occurance of high $^{29}$Si/$^{30}$Si ratios in many X grains may be explained.\\

\indent Figure \ref{rparam} is calculated at the specific shock time $t_0$ = 3$\times$10$^9$s. At 
that late time, the total number of Si atoms impacted is near 10$^{10}$, much 
less than the 10$^{12}$ Si atoms in a 1$\,\mu$m$^2$ grain of SiC. Likewise, the number 
of heavier Si isotopes that could be implanted is insufficient to alter the 
larger numbers actually found in SiC X grains. We would thus conclude that 
for a shock at 100 yr the Si isotopes in SiC X grains cannot be 
significantly altered. For trace elements, however, such as Fe or Mo, the 
implantation may still dominate \citep{2002ApJ...578L..83C}. On the other hand, 
Figure \ref{dparam} shows that the numbers in the 1$\,\mu$m$^2$ column decrease with shock 
time as (3$\times$10$^9$s/$t_0$)$^2$, so that a reverse shock encountered much
earlier may present the SiC condensate with more Si isotopes than actually 
exist in the grain. A reverse shock at $t_0$ = 10 yr, as might occur from the 
reaction to the presupernova wind, provides so many Si encounters that the 
grain encounters its own mass well before leaving the Si-rich region. This 
provides the new form of mixing discussed in section \ref{rev2}\\
\indent Table \ref{densdectab} shows the total number of possible interactions
between selected gas atoms and a 1$\,\mu$m grain as the grain travels through the 
uniform-velocity shell. These numbers are bounded by the upper limits detailed in Table
\ref{statictab}. For a grain condensing at 2.7M$_{\odot}$, the number of possible Si 
implantations is very close to the upper limit of Table \ref{statictab} ($^{29}$Si only 
17\% less), while the isotopes found farther out in the ejecta have significantly less 
than their respective upper limits, shown in Table \ref{statictab}. 
Because the density of the shell decreases as $t^{-2}$, regions far from the 
condensation of the grain have little effect, while the region that the grain condenses
in has a much more pronounced effect. This influence by nearby regions is seen 
in Table \ref{densdectab}. For example, the number of possible $^{100}$Mo implantations
increases if the grain condenses at 2.9M$_{\odot}$ compared with 2.7M$_{\odot}$, owing to the 
higher concentration of $^{100}$Mo in that region. It is not practical to illustrate all of 
these ideas in this paper. The full versions of Tables \ref{statictab} and \ref{densdectab} 
are found as electronic appendices. 

\subsection{Ion Implantation} 

\indent The physical environment of the grains following the passage of
the reverse shocks may be unique in that the grain is impacted by heavy
ions of anomalous isotopic composition at kinetic energies of 1 - 2 keV
per nucleon. The rate of collisions is high, about 10$^8$$\,$s$^{-1}$ ion
collisions for a 1$\,\mu$m$^2$ grain at $t$ = 10$^7$s. The associated heating rate 5 - 10
erg s$^{-1}$ keeps the grain above 1000K but does not evaporate it. The
collision energy is roughly equal to that of a well studied natural phenomenon, the
solar-wind bombardment of small surface grains on the moon; but the
rate of collisions is much smaller on the moon and the lunar grains
remain cold.\\ 
\indent Profiles of ion implantation have been measured for the lunar
soils \citep{1978LPSC....9.1655K,1978LPSC....9.1667Z}. \citeauthor{1978LPSC....9.1655K} 
describe stepwise heating release from individual olivine crystals showing ``bimodal
release patterns of Ne and Ar atoms, corresponding to a saturated
highly radiation damaged 30 nm surface layer and a less damaged zone
underneath which is populated by range straggling of solar wind.'' Their
Figure 5 shows ranges calculated by Lindhard theory for 1$\,$keV amu$^{-1}$ 
ions of He, Ne and Ar indicating that concentrations 1/10 of maximum are at
depths 60$\,$nm for Ar, 45$\,$nm for Ne and 30$\,$nm for He. This depth
approximately doubles for speeds of 2$\,$kev amu$^{-1}$. At the same time the
lunar grains are sputtered away slowly; and this will occur in the
supernova as well, with O atoms rather than He being the major
sputterers.\\ 
\indent In the simplest picture, refractory gas atoms in the supernova
(C, Si, Ti, Fe, etc.) will be implanted into a skin perhaps 50 - 100$\,$nm
thick and that skin will subsequently be sputtered away (primarily by O
impacts). The grain core might be thought to remain isotopically
unaffected in that picture. However, we point to three aspects of the
supernova problem that increase the diffusivity of supernova implanted
ions in comparison with the lunar case; namely, mean temperature above 1000$\,$K
and both temperature spikes and radiation damage by intense cosmic-ray
acceleration between the forward and reverse shocks. Higher $T$ increases
diffusivity in equilibrium crystals.  Accelerated particles burst
through the entire grain, leaving chemical excitations in their wake
and local thermal spikes that may enhance diffusion. We suggest that
detailed study of these effects are necessary before concluding that
ion implantation cannot alter the grain core. Furthermore, it will be
necessary to consider non-thermal events that the grain experiences
during its perhaps 10$^9$yr residence in the ISM before incorporation
into the meteorite. Although arguments that the grain cores will not
contain implanted ions must be addressed, it is equally necessary that
future work consider imaginatively the total integrated histories of
the presolar grains.

\subsection{Grain Aggregates}\label{agg}

\indent In general, grains condensed farther out in mass coordinate will
be moving faster than those condensed more centrally. Therefore grains
will not routinely overtake other grains. We note two exceptions to that
general expectation. Firstly, turbulent gases will cause grains to
collide, perhaps even destructively. Instabilities in the thin dense shell may cause 
differential grain speeds. We will not address this further,
although it may be more important than the second reason. Namely, small
grains can be overtaken by larger ones formed interior to them because
the small grains decelerate more rapidly than the larger ones. This must
certainly occur with or without turbulence.\\
\indent If the homologous expansion at the time of arrival of a reverse
shock at radius r is given by $v = H(t) r$ for matter within the shock. As
shown in Figure \ref{fig3}, grains existing at $r - \Delta r$ at the same time $t$ will,
after the shock reaches $r - \Delta r$, initially drift more slowly through
the shocked and decelerated gas by an amount
\begin{equation}
		\Delta v_g = -H(t) \Delta r
\end{equation}
Each grain loses drift speed at a rate given approximately by
\begin{equation}
		m \frac{dv_g}{dt} = -K a^2
\end{equation}
where a is the radius of the grain. This formula assumes that the
momentum loss by a fast grain is proportional to the rate at which its
area impacts slower gas atoms. Since $m = \rho\, a^3$, the simplest
expectation becomes $dv_g/dt = -K^{\prime}/a$; for example, a 0.1 micron grain
loses drift speed ten times faster than a 1 micron grain. This is a general expectation.\\
\indent The result is that larger grains continuously overtake smaller
ones. They collide at relative speeds that are much less than their
actual drift speeds with respect to the gas. We anticipate that hot
grains can stick together at these low relative speeds, thereby growing
grain aggregates, although we can present no evidence to support this
expectation. Visual inspection of many SiC X grains (such as in Figure \ref{SiCX}) show
features that look like grain assemblages; but this may simply be a property of
the SiC crystals themselves. Isotopic variations, if detected, could provide
such evidence. It is our hope that the new nanoSIMS technology \citep[for an overview of nanonSIMS, see][]{1999M&PSA..34..111S}
will enable the study of isotopic evidence from distinct subregions of a grain. 
For SiC, examination of Figure \ref{fig8} shows the sense of this to
be that large $^{28}$Si-rich grains overtake smaller grains having higher
$^{29,30}$Si fractions. In fact, this aggregation idea predicts such subgrains much richer in 
$^{29,30}$Si. This presents another way for grain aggregates to
acquire their $^{29,30}$Si content, despite many questions that must be
addressed. This rate of $^{29,30}$Si enrichment may even be more effective
than the direct implantation from the $^{29,30}$Si-rich gas. And either
mechanism suggests a possible reason for SiC X grains to typically have
about half the solar fractions of the heavy Si isotopes. 

\section{Discussion of Other Isotopes}

        Although SiC grains are built from Si and C atoms, isotopic
compositions of trace elements have been significant in defining the
X-grain class and in attributing them to supernovae \citep{2001ApJ...559..463A}
We here comment on four elements, N, Al, Mo, and Fe. for which good data
exist and for which the consequences of reverse shocks may be
significant. These illustrate both successes and puzzles of our model, as well as 
uses of Tables \ref{statictab} and \ref{densdectab}, which may be of interest 
to a wide range of other elements, eventually. 

\subsection{Nitrogen}

        Table \ref{densdectab} shows that all large SiC condensates, if initially condensed
from the range $2.7 < m < 3.3$ as we have argued, will implant $^{14}$N/$^{15}$N
isotopic ratios between 7 and 9, much smaller than the terrestrial
value, 272. In other words, the implanted atoms will be $^{15}$N-rich. The
$^{15}$N richness of the SiC X grains has been taken as an indication of
their supernova origin (strongly contrasted with N in mainstream SiC,
for example), and even as a diagnostic for X-grain classification
\citep{2001ApJ...559..463A}. But the measured X-grain ratios are not this
$^{15}$N-rich, lying mostly in the range 20 to 100, with 50 a typical value.
The problem within X grains, therefore, is not "why are they $^{15}$N-rich",
but "where do they get so much $^{14}$N".  Although ion implantation is an
attractive possibility, Table \ref{densdectab} shows that it is not an adequate source
of $^{14}$N. Our model therefore succeeds in explaining the $^{15}$N richness,
but we must seek other physical grounds for the high $^{14}$N content.
Although it goes far beyond the aims of this work, we point out that
the grains still move fast as they leave the contact discontinuity in
the absence of an early shock from the presupernova wind. Implantation
of $^{14}$N-rich atoms will continue as the grain slows down and even
afterward, when interstellar shock waves drive suprathermal $^{14}$N ions
into all interstellar grains.  We thus suggest that the $^{14}$N in the X
grains may arise primarily from ISM collisions. But it must also be
considered that small $^{14}$N-rich Si$_3$N$_4$ grains, or even graphite, may have
formed in the $^{14}$N-rich helium shell and may have been overtaken and
gathered by the faster moving (after deceleration) large SiC X grains
from the core. In such an interpretation, the large $^{15}$N-rich SuNoCon
may overtake and aggregate $^{14}$N-rich smaller grains that initially
condensed at greater radii. The new degree of freedom occurs if the
smaller grains that condensed further out condensed N more efficiently,
or held onto it more efficiently in the face of sputtering, than the
large $^{15}$N-rich SuNoCons from the basic SiC condensation zone. The N may
condense as refractory AlN or Si$_3$N$_4$ within these smaller, overtaken
grains, for example. In either case, the natural explantion of N
isotopes in X grains may be within reach.
 
\subsection{Aluminum}

        The case of $^{26}$Al has been especially vexing for X-grain
interpretation. One of the defining properties identifying the X
grain as a SuNoCon is the large isotopic ratio $^{26}$Al/$^{27}$Al that most
contained at the time they solidified, or as we would extend to in the
context of this paper, at the time they ceased to take on further
supernova atoms. This ratio is measured by the large excess of $^{26}$Mg in
these grains. It is usually the dominant Mg isotope, because SiC
condenses Al much more favorably than it does Mg. Thus the nearly
monoisotopic $^{26}$Mg was actually $^{26}$Al, and the measured initial $^{26}$Al/$^{27}$Al
ratios find almost half are greater that 0.1.  The observed $^{26}$Al/$^{27}$Al
initial ratios within the X grain are rather large for the SiC growth
zones tabulated in Table \ref{densdectab}, which are 3.5$\times$10$^{-3}$ for the implanted-ion
ratio. Although that ratio is suitable for some measured X grains,
others carry values up to 0.6.  Furthermore, Al is able to condense at
high T within SiC, so that it is not neccessary that the implanted ions
dominate the Al budget. An X grain may carry 10$^9$ - 10$^{10}$ Al atoms,
comparable to the number implanted in Table \ref{densdectab} for a shock at 10$^9$s.
This Al-isotope problem has been severe enough to have caused 
\citet{1997ApJ...486..824C} to show that the explosive He cap on a certain class of
Type Ia supernova is superior in naturally yielding such a large
$^{26}$Al/$^{27}$Al ratio. The large $^{27}$Al/$^{27}$Al ratios in the Type II supernova model
occur in the He zone and in the H-burning shell, far above the
condensation zone for SiC. We call to mind the N discussion above to
note that AlN condensed in that material will be both $^{26}$Al-rch and
$^{14}$N-rich, suggesting that assimilation of these small AlN grains by the
faster moving large SiC grain could, in principle, provide both the needed
$^{14}$N and $^{26}$Al. But this speculation too exceeds what we have shown.

\subsection{Molybdenum}

\begin{figure}[ht]
\centerline{\psfig{figure=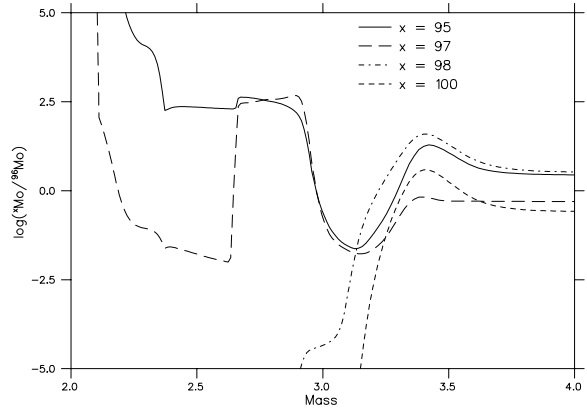,width=3.0in,angle=90}}
\caption{\it Mass fractions of selected Mo isotopes normalized to the
mass fraction of $^{96}$Mo, plotted versus mass coordinate. Only the region 
between $2.6 < m < 3.0$ has a marked overabundance of $^{95}$Mo and $^{97}$Mo.}\label{fig13} 
\end{figure}

        The case of molybdenum came to the fore as the result of new
ion-probe techniques called CHARISMA \citep*{2000LPI....31.1917P}. Four of seven
X grains studied in this way contained an initially surprising excesses
of $^{95}$Mo and $^{97}$Mo. Since \citet*{2000ApJ...540L..49M} had shown that this
unexpected isotopic richness of odd-A Mo isotopes was the result of
rapid neutron captures in the element Zr, an event that called for local
bursts of neutrons in the burning shells at the time of passge of the
forward shock, it became puzzling why so many SiC X grains would have
condensed in such burning shells. The reverse shocks add new dimension
to this \citep{2002ApJ...578L..83C}. One can suppose that the moving SiC
grains implant Mo atoms, and that therefore the frequency of occurance of these 
odd-A excesses can be high. First consider that the Mo atoms may
have been implanted. Table \ref{densdectab} shows that although $^{95}$Mo but not $^{97}$Mo are
somewhat enriched (measured relative to $^{96}$Mo) in the implanted column,
so too is $^{98}$Mo but not $^{100}$Mo. X grains measured by \citet{2000LPI....31.1917P}
have suggestively similar features, but also differences. Grain 209-1
has the largest excesses relative to $^{96}$Mo, namely 83\% for $^{95}$Mo and 70\%
for $^{97}$Mo, but also carries a 30\% excess of $^{98}$Mo and no excess of $^{100}$Mo.
Detailed analysis of this example also takes us beyond the goals of this paper.
What we now point out is that a significant fraction of the Mo isotopes
may also have been condensed rather than implanted. Figure \ref{fig13} shows
isotopic abundance ratios to $^{96}$Mo as a function of mass zone in the SiC
condensation zone. The region between $2.7 < m < 3.0$ is
unique in having large overabundances of only $^{95}$Mo and $^{97}$Mo. That this
is in the favored condensation zone suggests that condensation of Mo
along with the SiC was also important to the Mo budget in the X grains.
There does also exist another region of $^{95,97}$Mo excess in the He
burning shell (at $m = 7$ in the model used here); but $^{98}$Mo is also
abundant in that shell. Our conclusions from this limited discussion
would then be that the Mo atoms may all have been implanted, but that
for grains rich in $^{95,97}$Mo but not in $^{98}$Mo (if such exist) the
condensation component for Mo is primarily responsible.

\subsection{Iron}

        The excesses of $^{58}$Fe in SiC X grains provided \citet{2002ApJ...578L..83C}
with evidence in support of the importance of reverse shocks and
implantation of Fe atoms after that shock. Table \ref{densdectab} corroborates this
numerically. The number of $^{58}$Fe atoms encountered after the shock at
10$^9$s is from 1.4 to 1.8 times greater than the number of $^{56}$Fe atoms
encountered, for all condensation zones listed. The total number of Fe
atoms in an X grain may be 10$^8$ - 10$^9$, comparable to the number of $^{58}$Fe
atoms encountered in Table \ref{densdectab}. This means that the
implanted component can account only for $^{58}$Fe, and that most of the
remaining Fe atoms must instead have condensed within the SiC. A large plateau
of $^{54}$Fe excess lies just barely inside the SiC condensation zone $2.7 < m
< 3.2$, and even at $m = 2.7$ it remains the most overabundant Fe isotope. It
follows that $^{54}$Fe overabundance in X grains comes about as a result of
condensation near the inner boundary of the SiC condensation zone or by
virtue of significant instability and mixing between $m = 2.5$ and $m= 3.0$.\\
\indent Tables \ref{statictab} and \ref{densdectab} may also be used to estimate the numbers of Fe
atoms that are available for condensation as might occur if the reverse
shock arrived early enough that the growing SiC is stopped before leaving
the core. For a reverse shock at 3 yr, as could occur from a large
presupernova mass loss, the ion columns are 100 times
greater than those listed in Table \ref{statictab}, in which case a SiC that began
condensing at $m = 2.7$ would move forward to perhaps $m = 3.3$, by which
time the encountered Si column would equal the mass of the grain. The
composition of Fe encountered during that scenario would then
approximately equal the difference between the column above $m = 2.7$ and
the column above $m = 3.3$. That difference in Table \ref{statictab} is clearly also
$^{58}$Fe-rich. By this we intend only to illustrate the possible uses of
Tables \ref{statictab} and \ref{densdectab}, which appear much more completely in the electronic
tables.

\section{Summary and Conclusion}

\indent Our effort to understand why and how X-type SiC presolar grains
condensed within supernova interiors prior to any mixing with
circumstellar matter has produced several new discoveries and several
hypotheses. They are:\\
\indent 1. Only where n(Si) $>$ n(C) can SiC condense because graphite
normally condenses prior to SiC and exhausts the C. In the absence of
specific kinetic condensation model for SiC, we take the thermal
equilibrium guidelines to suggest that n(Si) $>$ 10 n(C) is required.\\ 
\indent 2. In the new 25M$_{\odot}$ model by WHW, SiC condenses between 2.7
 $<$ m $<$ 3.2 because of conclusion 1. \\
\indent 3. Prior to condensation, a reverse shock is returned to the
core by the deceleration of the outward moving shock while it propogates
through the H envelope. That deceleration happens because of the radial
increase of the product $\rho$r$^3$ within the H envelope, and causes the
pileup of hot gas that sends the inward pressure wave. After about 10$^6$
s this reverse shock creates a high density shell between 2.7 $<$ m $<$
3.6, where the density is 10$^2$ greater than in neighboring mass zones.
This high density is maintained by momentum flux into that region,
which has higher pressure than the surroundings.\\
\indent 4. The mass zones  n(Si) $>$ 10 n(C) lie within the inner part of
that high-density shell, between 2.7 $<$ m $<$ 3.2, which is therefore
where SiC SuNoCons condense. At larger m graphite condenses instead of
SiC. \\
\indent 5. The excess O does not lock up the Si and C atoms owing to the
radioactive disruption of SiO and CO, allowing carbonaceous
condensation in oxidizing gas \citep{1999Sci...283.1290C}.\\
\indent 6. The $^{28}$Si-richness of X grains of SiC is explained by 
the isotopic constitution of the SiC condensing shell 2.7 $<$ m $<$ 3.2. We
also suggest that the local nucleosynthesis in this shell in the presolar
supernovae that parented the X grains produced silicon that is
isotopically lighter than in the WHW calculation for solar metallicity
by a factor two owing to the secondary-nucleosynthesis properties of
the heavy isotopes of Si \citep{1996ApJ...472..723T}. These factors
conspire to explain the absence of supernova SiC SuNoCons having
isotopically heavy Si.\\
\indent 7. Between 10$^7$s and 10$^9$s a second reverse shock moves inward
through the mass of the core. It is launched by the collision of the
outward shock with the presupernova wind \citep{1994ApJ...420..268C}.
In the WHW 25$_{\odot}$ model the circumstellar wind is 12M$_{\odot}$ at the time
of the explosion. This shock decelerates the gas to about 60\% of its
preshock velocity \citep{1999ApJS..120..299T}.\\
\indent 8. Following the preceeding reverse shock, the already
condensed SuNoCons move forward rapidly through the declerated
gas. Their radial speed is equal to their preshock speed. Thus the
grains move forward through the overlying gas at a relative speed that
is initially near 40\% of their preshock speed, roughly 500-600 km s$^{-1}$ for
matter near m = 3 in the 25M$_{\odot}$ WHW model.\\
\indent 9. The forward moving grains implant struck atoms to depths
near 0.05 $\mu$m, causing change in the isotopic composition of the
SuNoCon \citep{2002ApJ...578L..83C}. The 500 km s$^{-1}$ collisions also cause
sputering of the grains, primarily by O atoms.\\
\indent 10. Owing to the overlying column, the SuNoCons decelerate with
respect to the gas until they move with the gas. Because the SuNoCons
remain cool, about 1000K, condensation of hot atoms (10$^6$K) from the
local gas continues. This causes mixing of the isotopic composition of
the condensate. It is a new physical environment for condensation.\\ 
\indent 11. As grains decelerate, large grains overtake smaller hot
grains at low relative speeds, perhaps allowing them to coallesce. This
also causes isotopic mixing of the evolving SuNoCon. Both 10 and 11 may
explain  when the heavy Si isotopes and the $^{58}$Fe isotope are added to
the SuNoCon \citep{2002ApJ...578L..83C}. This may also explain why these grains 
often look like the assemblages of smaller grains (Figure \ref{SiCX}).\\
\indent 12. We calculate and tabulate the numbers of atoms in the
column overlying a SuNoCon that is already condensed at mass
coordinate r$_0$ at time t$_0$ when the reverse shock arrives. Several
comsochemical applications are illustrated by the composition of this
column.\\
\indent 13. We calculate the numbers of atoms in the overlying column of
12 also taking into account the continuing expansion of the supernova
while the grain propagates though that column.\\
\indent 14. Between 10$^9$s and 10$^{11}$s, a third reverse shock propogates
into the supernova mass owing to collision of the outward shock with
the external ISM. Differential motion of grains and gas occurs again,
allowing further sputtering and ion implnation. The altered SuNoCon
emerges from the contact discontinuity into the ISM.\\
\indent Each of these topics raises many additional questions that have
not yet been answered. We discussed many of these: the need for 3-D
hydrodynamic simulations; the need to study simultaneous turbulent and
molecular diffusion; whether the isotopicaly anomalous skin of a 
SuNoCon can be homogenized within its bulk; detailed motion of grains relative to gas; 
whether hot grains coalesce in
low speed collisions; the origin of TiC subgrains within the SiC; the
actual kinetic description of SiC condensation. These many uncertainties
and new questions notwithstanding, we have presented a physical road map
to the existence and properties of presolar SiC grains from supernovae,
and how they contain information about the young supernova remnant.\\

\acknowledgments

Research by ED and DDC is supported by the National Aeronautics and Space
Administration under Grant NAG5-11871 issued through the Office of Space
Science

AH is supported in part by the Department of Energy under grant B341495 to the 
Center for Astrophysical Thermonuclear Flashes at the
University of Chicago and acknowledges support by a Fermi Fellowship
of the Enrico Fermi Institute at The University of Chicago.

\begin{deluxetable}{ccc}
\tabletypesize{\scriptsize}
\tablecaption{($^i$Si/$^{28}$Si)/($^i$Si/$^{28}$Si)$_{\odot}$ From a Half-Solar Metallicity 25M$_{\odot}$ Star\label{massratiotab}}
\tablewidth{0pt}
\tablehead{
\colhead{$\Delta$m} & \colhead{$^{29}$Si/$^{28}$Si}   & \colhead{$^{30}$Si/$^{28}$Si}
}
\startdata
2.7 - 3.0M$_{\odot}$ &0.18 &0.51 \\
2.7 - 3.2M$_{\odot}$ &0.29 &0.76 \\
2.7 - 3.6M$_{\odot}$ &0.46 &0.95 \\
\enddata
\end{deluxetable}

\begin{deluxetable}{ccccc}
\tabletypesize{\scriptsize}
\tablecaption{Number of Atoms in a Static Column at 10$^9$s for Selected Isotopes \tablenotemark{a}\label{statictab}}
\tablewidth{0pt}
\tablehead{
\colhead{} & \multicolumn{4}{c}{Grain Condenses At} \\
\cline{2-5} \\
\colhead{Isotope} & \colhead{2.7M$_{\odot}$}   & \colhead{2.9M$_{\odot}$}   &
\colhead{3.1M$_{\odot}$} & \colhead{3.3M$_{\odot}$}
}
\startdata
$^{14}$N &1.366$\times$10$^{10}$ &1.366$\times$10$^{10}$ &1.366$\times$10$^{10}$ &1.366$\times$10$^{10}$ \\ 
$^{15}$N &1.562$\times$10$^{8}$ &1.503$\times$10$^{8}$ &1.436$\times$10$^{8}$ &1.374$\times$10$^{8}$ \\
$^{26}$Al &6.925$\times$10$^{7}$ &6.919$\times$10$^{7}$ &6.842$\times$10$^{7}$ &5.860$\times$10$^{7}$ \\
$^{27}$Al &2.269$\times$10$^{10}$ &2.235$\times$10$^{10}$ &2.105$\times$10$^{10}$ &1.898$\times$10$^{10}$ \\
$^{28}$Si &1.197$\times$10$^{11}$ &6.879$\times$10$^{10}$ &3.630$\times$10$^{10}$ &2.042$\times$10$^{10}$ \\
$^{29}$Si &9.646$\times$10$^{9}$ &9.040$\times$10$^{9}$ &7.612$\times$10$^{9}$ &5.772$\times$10$^{9}$ \\
$^{30}$Si &9.894$\times$10$^{9}$ &8.714$\times$10$^{9}$ &6.197$\times$10$^{9}$ &3.628$\times$10$^{9}$ \\ 
$^{54}$Fe &4.495$\times$10$^{7}$ &4.305$\times$10$^{7}$ &4.258$\times$10$^{7}$ &4.253$\times$10$^{7}$ \\
$^{56}$Fe &8.163$\times$10$^{8}$ &7.953$\times$10$^{8}$ &7.866$\times$10$^{8}$ &7.845$\times$10$^{8}$ \\
$^{57}$Fe &7.347$\times$10$^{7}$ &7.329$\times$10$^{7}$ &7.269$\times$10$^{7}$ &7.145$\times$10$^{7}$ \\
$^{58}$Fe &3.035$\times$10$^{8}$ &3.022$\times$10$^{8}$ &2.951$\times$10$^{8}$ &2.808$\times$10$^{8}$ \\
$^{92}$Mo &3.647$\times$10$^{3}$ &1.439$\times$10$^{3}$ &2.887$\times$10$^{2}$ &2.878$\times$10$^{2}$ \\
$^{94}$Mo &2.579$\times$10$^{3}$ &2.539$\times$10$^{3}$ &5.334$\times$10$^{2}$ &2.431$\times$10$^{2}$ \\
$^{95}$Mo &2.910$\times$10$^{4}$ &2.909$\times$10$^{4}$ &2.905$\times$10$^{4}$ &2.885$\times$10$^{4}$ \\
$^{96}$Mo &1.378$\times$10$^{4}$ &1.378$\times$10$^{4}$ &1.315$\times$10$^{4}$ &1.076$\times$10$^{4}$ \\
$^{97}$Mo &6.540$\times$10$^{3}$ &6.525$\times$10$^{3}$ &6.496$\times$10$^{3}$ &6.433$\times$10$^{3}$ \\
$^{98}$Mo &3.524$\times$10$^{4}$ &3.524$\times$10$^{4}$ &3.524$\times$10$^{4}$ &3.359$\times$10$^{4}$ \\
$^{100}$Mo &2.937$\times$10$^{3}$ &2.937$\times$10$^{3}$ &2.937$\times$10$^{3}$ &2.896$\times$10$^{3}$ \\
\enddata


\tablenotetext{a}{These quantities signify the upper limit on the number of possible
interactions of the isotope with the grain. Number of atoms in 1 $\mu$m$^2$ static column
following a reverse shock at 10$^9$s. Numbers scale as (10$^9$s/t$_0$)$^2$ }

\end{deluxetable}

\begin{deluxetable}{ccccc}
\tabletypesize{\scriptsize}
\tablecaption{Number of Atoms in a Moving Column \tablenotemark{a} for a Shock Time 10$^9$s for Selected Isotopes \tablenotemark{b}\label{densdectab}}
\tablewidth{0pt}
\tablehead{
\colhead{} & \multicolumn{4}{c}{Grain Condenses At} \\
\cline{2-5} \\
\colhead{Isotope} & \colhead{2.7M$_{\odot}$}   & \colhead{2.9M$_{\odot}$}   &
\colhead{3.1M$_{\odot}$} & \colhead{3.3M$_{\odot}$}
}
\startdata
$^{14}$N &6.105$\times$10$^{8}$ &6.109$\times$10$^{8}$ &6.111$\times$10$^{8}$ &6.113$\times$10$^{8}$ \\
$^{15}$N &8.584$\times$10$^{7}$ &8.023$\times$10$^{7}$ &7.370$\times$10$^{7}$ &6.763$\times$10$^{7}$ \\
$^{26}$Al &5.336$\times$10$^{7}$ &5.347$\times$10$^{7}$ &5.280$\times$10$^{7}$ &4.308$\times$10$^{7}$ \\
$^{27}$Al &1.577$\times$10$^{10}$ &1.549$\times$10$^{10}$ &1.422$\times$10$^{10}$ &1.217$\times$10$^{10}$ \\
$^{28}$Si &1.128$\times$10$^{11}$ &6.226$\times$10$^{10}$ &2.986$\times$10$^{10}$ &1.403$\times$10$^{10}$ \\
$^{29}$Si &8.029$\times$10$^{9}$ &7.449$\times$10$^{9}$ &6.035$\times$10$^{9}$ &4.204$\times$10$^{9}$ \\
$^{30}$Si &8.855$\times$10$^{9}$ &7.703$\times$10$^{9}$ &5.199$\times$10$^{9}$ &2.639$\times$10$^{9}$ \\
$^{54}$Fe &4.829$\times$10$^{6}$ &2.939$\times$10$^{6}$ &2.469$\times$10$^{6}$ &2.426$\times$10$^{6}$ \\
$^{56}$Fe &1.091$\times$10$^{8}$ &8.842$\times$10$^{7}$ &7.989$\times$10$^{7}$ &7.789$\times$10$^{7}$ \\
$^{57}$Fe &2.293$\times$10$^{7}$ &2.281$\times$10$^{7}$ &2.224$\times$10$^{7}$ &2.104$\times$10$^{7}$ \\
$^{58}$Fe &1.540$\times$10$^{8}$ &1.531$\times$10$^{8}$ &1.463$\times$10$^{8}$ &1.322$\times$10$^{8}$ \\
$^{92}$Mo &3.365$\times$10$^{3}$ &1.165$\times$10$^{3}$ &1.628$\times$10$^{1}$ &1.534$\times$10$^{1}$ \\
$^{94}$Mo &2.353$\times$10$^{3}$ &2.321$\times$10$^{3}$ &3.183$\times$10$^{2}$ &2.818$\times$10$^{1}$ \\
$^{95}$Mo &1.531$\times$10$^{4}$ &1.534$\times$10$^{4}$ &1.534$\times$10$^{4}$ &1.516$\times$10$^{4}$ \\
$^{96}$Mo &7.832$\times$10$^{3}$ &7.855$\times$10$^{3}$ &7.240$\times$10$^{3}$ &4.867$\times$10$^{3}$ \\
$^{97}$Mo &2.614$\times$10$^{3}$ &2.606$\times$10$^{3}$ &2.581$\times$10$^{3}$ &2.521$\times$10$^{3}$ \\
$^{98}$Mo &2.181$\times$10$^{4}$ &2.188$\times$10$^{4}$ &2.192$\times$10$^{4}$ &2.031$\times$10$^{4}$ \\
$^{100}$Mo &1.658$\times$10$^{3}$ &1.663$\times$10$^{3}$ &1.665$\times$10$^{3}$ &1.627$\times$10$^{3}$ \\
\enddata


\tablenotetext{a}{Column limits are from condensation m to the contact discontinuity.}
\tablenotetext{b}{Number of atoms per 1$\mu$m$^2$, Distance parameter, t$_0$$\Delta$V = 5$\times$10$^{16}$cm}
\tablecomments{The full, machine readable version of this table is availible in the paper's electronic supplement.}
\end{deluxetable}

\bibliographystyle{apj}



\end{document}